\documentstyle[preprint,aps,floats,epsfig]{revtex}

\begin{document}
\preprint{\vbox{\hbox{}\hbox{}\hbox{}}}
\draft
\title {A study of semi-inclusive charmless $B \to \pi X$ decays}
\author{C.~ S.~ Kim\footnote{cskim@mail.yonsei.ac.kr,~~~
http://phya.yonsei.ac.kr/\~{}cskim/},~~ Jake~ Lee~~ and~~ Sechul~
Oh\footnote{scoh@phya.yonsei.ac.kr}}
\address{Department of Physics and IPAP, Yonsei University, Seoul,
120-749, Korea}
\author{Jong Sung Hong$^{a}$, Deuk Young Kim$^{b}$, and Hyung Sang Kim$^{c}$}
\address{a: Dept. of General Education, Samchok Univ., Kangwondo,
Korea \\
b: Dept. of Semiconductor Science, Dongguk Univ., Seoul, Korea \\
c: Dept. of Physics, Dongguk Univ., Seoul, Korea}
\maketitle
\begin{abstract}

\noindent We study semi-inclusive charmless decays $B \to \pi X$
in detail, such as $\overline B^0 \to \pi^{\pm (0)} X$,
$B^0 \to \pi^{\pm (0)} X$, $B^{\pm} \to \pi^{\pm (0)} X$,
where $X$ does not contain a charm (anti)quark.
We find that the process $\overline B^0 \to \pi^- X$ ($B^0 \to \pi^+ X$)
can be particularly
useful for determination of the CKM matrix element $|V_{ub}|$.
We calculate and present the branching ratio (BR) of $\overline B^0 \to \pi^- X$
as a function of $|V_{ub}|$, with an estimate of possible
uncertainties.
It is expected that the BR is an order of $10^{-4}$.
Our estimation indicates that one can phenomenologically determine $|V_{ub}|$
with reasonable accuracy by measuring the BR of $\overline B^0 \to \pi^- X$
($B^0 \to \pi^+ X$).
\end {abstract}

\newpage
\section{Introduction}

The source of CP violation in the Standard Model (SM) with three
generations is a phase in the
Cabibbo-Kobayashi-Maskawa (CKM) matrix \cite{ckm}.
A precise measurement of the CKM matrix elements is one of the key issues
in the study of $B-$mesons and $B-$factory experiments.
In particular, the accurate determination of $V_{ub}$ is one of the most
challenging problems in $B$ physics.
Its non-vanishing value is a necessary condition for CP violation to occur
in the SM and its accurate value can put strong constraints even on the unitarity
triangle: for instance, on the magnitude of the CP violating phase
$\beta (\equiv \phi_1)$.

Theoretical and experimental studies for probing $V_{ub}$ have been mostly
focused on the semileptonic $B-$meson decays.  The present best experimental
data for $V_{ub}$ come from measurements of
the exclusive decay $B \to \rho l \bar \nu$  and the inclusive decay
$B \to X_u l \bar \nu$,
but these measurements suffer from large uncertainties due to
model-dependence and other theoretical errors.  For example, the CLEO
result obtained using the exclusive semileptonic decay
$B \to \rho l \bar \nu$ \cite{cleo} :
\begin{eqnarray}
  |V_{ub}| = ( 3.25 \pm 0.14(\rm stat.)^{+0.21}_{-0.29} (\rm syst.)
  \pm 0.55 (\rm model) ) \times 10^{-3}~.
\end{eqnarray}
The OPAL data obtained using the inclusive decay $B \to X_u l \bar \nu$
\cite{opal} :
\begin{eqnarray}
  |V_{ub}| = ( 4.00 \pm 0.65(\rm stat.)^{+0.67}_{-0.76} (\rm syst.)
  \pm 0.19 (\rm HQE) ) \times 10^{-3}~.
\end{eqnarray}
The method using the exclusive semileptonic decays involves hadronic form
factors, such as $F^{B \to \pi}$ or $A^{B \to \rho}$, whose values are heavily
model-dependent and cause large uncertainties.
The difficulty of using the inclusive charmless decay $B \to X_u l \bar \nu$ is
in discriminating this process from the dominant background $B \to X_c l \nu$
decay \cite{bkp}, whose branching ratio is more than 50 times larger than that of
$B \to X_u l \bar \nu$.

Although traditional difficulties with the understanding of non-leptonic $B$
decays have prevented their use in determination of the CKM matrix elements,
the possibility of measuring $|V_{ub}|$ via non-leptonic decays of $B-$mesons
to exclusive or inclusive final states has been also theoretically explored
\cite{koide,choud,kkln,fa}.

In this work we study semi-inclusive charmless decays $B \to \pi X$ and
investigate the possibility of extracting $|V_{ub}|$ from these processes.
Compared to the exclusive decays, these semi-inclusive decays are generally
expected to have less hadronic uncertainties and larger branching ratios.
There are several possible processes in $B \to \pi X$ type decays, such as
$\overline B^0 \to \pi^{\pm (0)} X$, $B^0 \to \pi^{\pm (0)} X$,
$B^{\pm} \to \pi^{\pm (0)} X$, where $X$ does not contain a charm
(anti)quark.
The class of nonleptonic two-body decay modes and their advantages
within general arguments were previously discussed in Ref.
\cite{soni}, and semi-inclusive two-body decays within the QCD factorization were
also studied in Ref. \cite{cheng}.

In Sec. II we first classify all those $B \to \pi X$ processes,
and we identify a certain mode,
$\overline B^0 \to \pi^- X$, whose analysis is theoretically clean and which
can be used for determining $|V_{ub}|$.  Then, in Sec. III we study the mode
$\overline B^0 \to \pi^- X$ in detail and propose a method to extract $|V_{ub}|$.
That is, we calculate the branching ratio (BR) of
this mode using the full effective Hamiltonian in the framework of the
generalized factorization, and present the result as a function of
$|V_{ub}|$ with an estimation of possible uncertainties.   We also consider
the $B^0 - \overline B^0$ mixing effect through $\overline B^0 \to B^0 \to \pi^- X$.
The conclusions are in Sec. IV.

\section{Classification of Semi-inclusive charmless $B \to \pi X$ decays}

Among the semi-inclusive charmless $B \to \pi X$ decays, let us first consider
the mode $\overline B^0 \to \pi^- X$.  Contributions for the decay amplitude of
this mode arise from the color-favored tree ($b \to u \bar u d$) diagram and
the $b \to d$ penguin diagram (see Fig. 1), and the tree diagram contribution
dominates.
\begin{figure}[htb]
\begin{center}
\epsfig{file=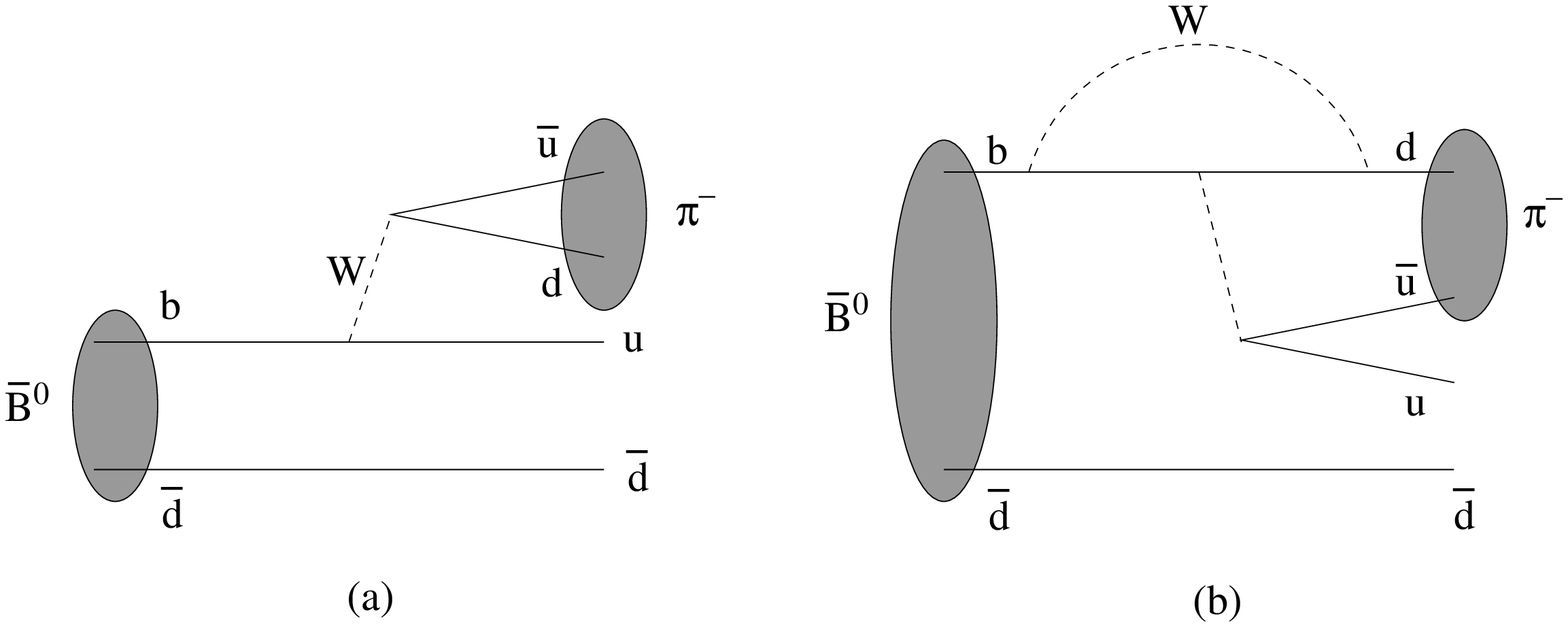,width=12cm}
\end{center}
\caption {Feynman diagrams of $\overline B^0 \to \pi^- X$ decay:
(a) the color-favored tree diagram,
and (b) the $b \to d$ penguin diagram. }
\vspace{0.5cm}
\end{figure}
The charged pion $\pi^-$ in the final state can be produced
via $W-$boson emission at tree level and is expected to be energetic
($E_{\pi^-} \sim m_B / 2$).  The decay amplitude can be approximated as
\begin{eqnarray}
A( \overline B^0 \to \pi^- X ) \simeq A (b \to \pi^- u) \cdot h(u \bar d \to X(u \bar d))
\approx A (b \to \pi^- u)~,
\end{eqnarray}
where $h$ denotes a hadronization function describing the combination of
the $u \bar d$ pair to make the final state $X$.  To obtain the decay rate,
$X(u \bar d)$ should be summed over all the possible states, such as $\pi^+ \pi^0$,
$\pi^+ \pi^+ \pi^-$ etc, so this process is effectively a \emph{two-body}
decay process of $b \to \pi^- u$ in the parton model approximation\footnote{
We notice that the dominant tree contribution (Fig. 1(a)) is  diagramatically
similar to the inclusive semileptonic decay, $b \to  [l^- \nu] u$,
and can be approximated to the free quark decay of $b \to \pi^- u$ within the HQET. }.
Thus, in this specific mode, no hadronic form factors (except the pion decay
constant $f_\pi$)
are involved, and as a result the model-dependence does not appear to be severe.
We note that the \emph{energetic} charged pion\footnote{
The net electric charge of $X$ should be \emph{positive} so that
such energetic $\pi^-$ cannot be produced from the inclusive
$X = \pi^+ \pi^0$, $\pi^+ \pi^+ \pi^-$, etc.}
$\pi^-$ in the final
state can be a characteristic signal for this mode.
We will show in
Fig. 3  the decay distribution, $d\Gamma \over dE_{\pi}$,
for $\overline B^0 \to \pi^- X$ as a function of the charged pion energy $E_{\pi}$.
(For a detailed explanation, see Sec. III.)
Like a two-body decay, a peak appears around $E_{\pi^-} = m_B /2$.
\begin{figure}[htb]
\begin{center}
\epsfig{file=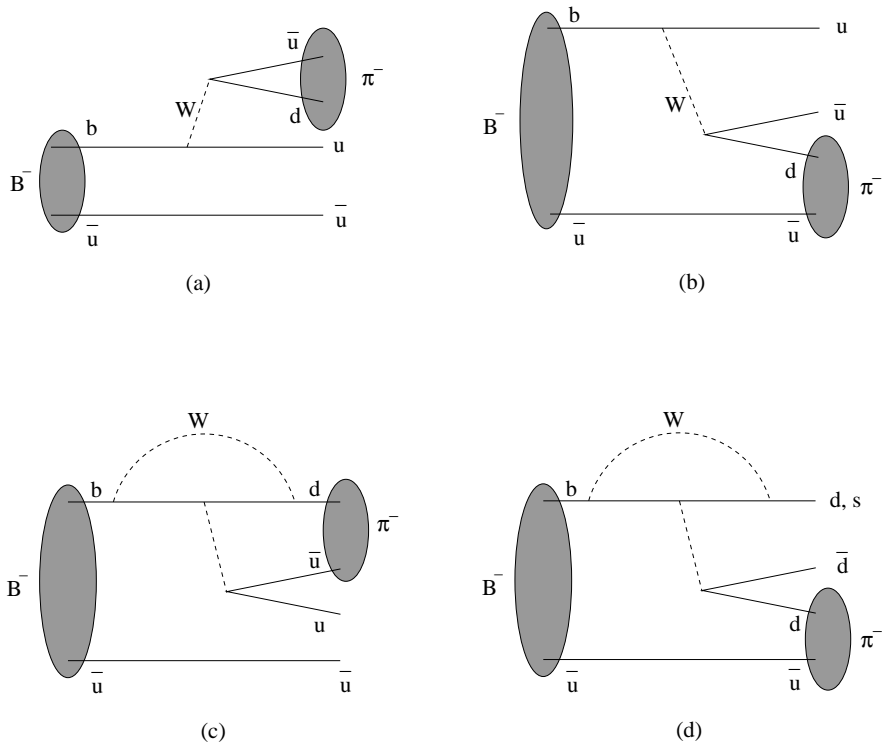,width=12cm}
\end{center}
\caption {Feynman diagrams of $B^- \to \pi^- X$ decay:
(a) the color-favored tree diagram,
(b) the color-suppressed tree diagram, (c) the $b \to d$ penguin diagram,
and (d) the $b \to d$ and $b \to s$ penguin diagram. }
\vspace{0.5cm}
\end{figure}

Now let us consider the mode $B^- \to \pi^- X$.
As shown in Fig. 2, various contributions are responsible for this process:
the color-favored tree diagram, the color-suppressed tree diagram, the $b \to d$
and $b \to s$ penguin diagrams.  The color-favored tree contribution (Fig. 2(a))
and  $b \to d$ penguin contributions (Fig. 2(c)) are
similar to those in $\overline B^0 \to \pi^- X$, which are effectively two-body type
($b \to \pi^- u$) processes.  However, the color-suppressed tree (Fig. 2(b)) and
other penguins (Fig. 2(d)) differ from those in $\overline B^0 \to \pi^- X$.
In fact, these two diagrams correspond to effectively a \emph{three-body} decay process
of $B^- \to \pi^- u \bar u$ (or $\pi^- d \bar d$)
in the parton model approximation, with the decay amplitude
\begin{eqnarray}
A( B^- \to \pi^- X ) \simeq A (B^- \to \pi^- u \bar u) \cdot
h(u  \bar u \to X(u \bar u))
\approx A (B^- \to \pi^- u \bar u)~.
\end{eqnarray}
In Figs. 2(b) and 2(d), the charged pion $\pi^-$ in the final state
contains the spectator antiquark $\bar u$. So the analysis involves
the hadronic form factor for the $B \to \pi$ transition which
is highly model-dependent.
Furthermore, the $b \to s$ penguin contribution (Fig. 2(d)) is not suppressed compared to
the tree contributions, but dominant in this mode.
The decay distribution, $d\Gamma \over dE_{\pi}$, for $B^- \to \pi^- X$
will be shown in Fig. 4, as a function of the charged pion energy $E_{\pi}$.
The three-body type contribution from the $b \to s$ penguin is the dominant one.
Therefore, compared to the case of $\overline B^0 \to \pi^- X$,
the analysis of this mode is much more complicated and involves
larger uncertainties.

Other modes of the type $B \to \pi X$ can be similarly classified.
For instance, in the mode $B^0 \to \pi^- X$, the color-favored tree
($\bar b \to \bar u u \bar d$ and $\bar b \to \bar u u \bar s$) diagrams and
the $b \to d$ and $b \to s$ penguin diagrams are responsible for the decay process.
In this case, the charged pion $\pi^-$ contains the spectator quark $d$ so that
the process is effectively a three-body decay $B^0 \to \pi^- u \bar d ~(\bar s)$
and the hadronic form factor for the $B \to \pi$ transition is involved.
Other processes are basically
a combination of the two-body decay process ($b \to \pi q$) and the three-body
decay process ($B^- \to \pi^- q \bar q^{\prime}$).

\section{Analysis of $\overline B^0 \to \pi^- X$ decay}

In the previous section, we have seen that the process $\overline B^0 \to \pi^- X$ is
particularly interesting, because it is effectively the two-body decay
process $b \to \pi^- u$ in the parton model approximation,
and no uncertainty from hadronic form factors is
involved.   Thus, its theoretical analysis is expected to be quite clean.

The relevant $\Delta B =1$ effective Hamiltonian for hadronic $B$ decays
can be written as
\begin{eqnarray}
H_{eff}^{q} &=& {G_F \over \sqrt{2}} \left[ V_{ub}V^{*}_{uq} (c_1 O^q_{1u}
+c_2 O^q_{2u}) + V_{cb}V^{*}_{cq} (c_1 O^q_{1c} +c_2 O^q_{2c}) \right.
\nonumber \\
&-& \left. \sum_{i=3}^{10} \left( V_{ub} V^*_{uq} c_i^u +V_{cb} V^*_{cq} c_i^c
+V_{tb} V^*_{tq} c_i^t \right) O_i^q \right]  \nonumber \\
&+& H.C.~ ,
\end{eqnarray}
where $O^q_i$'s are defined as
\begin{eqnarray}
O^q_{1f} &=& \bar q \gamma_{\mu} L f \bar f \gamma^{\mu} L b,  \ \
O^q_{2f} = \bar q_{\alpha} \gamma_{\mu} L f_{\beta} \bar f_{\beta} \gamma^{\mu}
L b_{\alpha}~,   \nonumber \\
O^q_{3(5)} &=& \bar q \gamma_{\mu} L b \sum_{q^{\prime}} \bar q^{\prime}
\gamma^{\mu} L(R) q^{\prime},  \ \
O^q_{4(6)} = \bar  q_{\alpha} \gamma_{\mu} L b_{\beta} \sum_{q^{\prime}}
\bar q^{\prime}_{\beta}  \gamma^{\mu}
L(R) q^{\prime}_{\alpha}~,  \nonumber \\
O^q_{7(9)} &=& {3 \over 2} \bar q \gamma_{\mu} L b \sum_{q^{\prime}}
e_{q^{\prime}} \bar q^{\prime}  \gamma^{\mu}
R(L) q^{\prime} , \ \
O^q_{8(10)} ={3 \over 2} \bar q_{\alpha} \gamma_{\mu} L b_{\beta}
\sum_{q^{\prime}}  e_{q^{\prime}} \bar
q^{\prime}_{\beta} \gamma^{\mu} R(L) q^{\prime}_{\alpha}~ ,
\end{eqnarray}
where $L(R) = (1 \mp \gamma_5)$, $f$ can be $u$ or $c$  quark, $q$ can be
$d$ or $s$ quark,  and $q^{\prime}$ is
summed over $u$, $d$, $s$, and $c$ quarks. $\alpha$ and  $\beta$ are  the
color indices.
$c_i$'s  are the Wilson  coefficients (WC's), and
we use the effective WC's for the process
$b \to d \bar qq'$ from Ref.\cite{desh}.  The regularization
scale is taken to be $\mu=m_b$. The operators
$O_1$, $O_2$ are the tree level and QCD corrected operators,  $O_{3-6}$ are
the gluon induced  strong penguin operators,
and finally  $O_{7-10}$ are the electroweak
penguin operators due to $\gamma$ and $Z$ exchange, and  the box diagrams at
loop level.

Now we calculate the decay amplitude for the semi-inclusive decay
$\overline B^0 \to \pi^- X$, where $X$ can contain an up quark and a down antiquark.
In the generalized factorization approximation, the decay amplitude is given by
\begin{eqnarray}
{\cal M} &=& \langle \pi^- X| H_{eff} | \overline B^0 \rangle ~, \\
&=& i{G_F \over \sqrt{2}} f_{\pi}
\langle X| \bar u \left[ r (1+\gamma^5)
+ l (1-\gamma^5) \right] b|\overline B^0 \rangle ~,
\end{eqnarray}
where we have defined the followings:
\begin{eqnarray}
r &=& m_b w_1~,  \nonumber \\
l &=& - m_u w_1 -{m_{\pi}^2 \over  m_d + m_u} w_2~, \nonumber \\
w_1 &=& V_{ub} V^*_{ud} \left( {c_1 \over N_c} + c_2 \right) +{A_3 \over N_c} +A_4
+{A_9 \over N_c} + A_{10}~,  \nonumber \\
w_2 &=& - 2 \left( {A_5 \over N_c} + A_6 +{A_7 \over N_c} +A_8 \right)~ .
\end{eqnarray}
Here $N_c$ denotes the effective number of color and
\begin{eqnarray}
A_i &=& - \sum_{q=u,c,t} V_{qb} V^*_{qd} c^q_i \ .
\end{eqnarray}
We have used the relations
\begin{eqnarray}
\langle \pi^-| \bar d \gamma^{\mu} \gamma^5 u|0 \rangle
&=& -i f_{\pi} p^{\mu}_{\pi}~, \\
\langle \pi^-| \bar d \gamma^5 u|0 \rangle
&=& -i {f_{\pi} m^2_{\pi} \over m_d + m_u}~,  \label{pion}
\end{eqnarray}
where $f_{\pi}$ and $p^{\mu}_{\pi}$ are the decay constant and the momentum of
pion, respectively, and $m_{\pi}~(m_i)$ is the mass of pion ($i$ quark).
In Eq. (\ref{pion}) the free quark equation of motion has been used.

Then,
\begin{eqnarray}
|{\cal M}|^2 &=& {G_F^2 \over 2} f_{\pi}^2 \sum_{X} \left| \langle X| \bar u
\left[ r (1+\gamma^5) + l (1-\gamma^5) \right] b| \overline B^0 \rangle
\right|^2 (2\pi)^4 \delta^4 (p_B -p_{\pi} -p_X)    \nonumber
\\ &=& {G_F^2 \over 2} f_{\pi}^2 \sum_{X} \langle \overline B^0| J |X \rangle
\langle X| J^{\dagger} |\overline B^0 \rangle (2\pi)^4 \delta^4 (p_B -p_{\pi} -p_X)~,
\end{eqnarray}
where
\begin{eqnarray}
J^{\dagger} = \bar u \left[ r (1+\gamma^5) +l (1-\gamma^5) \right] b~.
\end{eqnarray}
In the parton model approximation we take the leading order term in the product
of the above matrix elements, which corresponds to interpretation of the above
process as $b(p_b) \rightarrow \pi^-(p_{\pi}) +u(p_u)$ \cite{he}.
Then, $|{\cal M}|^2$ can be expressed as
\begin{eqnarray}
|{\cal M}|^2 &=& 2 G_F^2 f_{\pi}^2 \left[ (|r|^2 +|l|^2) (p_u \cdot p_b) +2
Re(r l^*) m_u m_b \right]  \nonumber \\
&=& |V_{ub}|^2 {\cal M}_2 + |V_{ub}| {\cal M}_1 + {\cal M}_0 ~,
\label{Msquared}
\end{eqnarray}
where
\begin{eqnarray}
{\cal M}_2 &=& 2 G_F^2 f_{\pi}^2 |V_{ud}^*|^2 \{ |c^{ut}|^2 {\cal X}
+ 4 |\tilde c^{ut}|^2 {\cal Y} +2 Re[(c^{ut}) (\tilde c^{ut})^* ] {\cal Z} \}~,
\nonumber \\
{\cal M}_1 &=& 2 G_F^2 f_{\pi}^2 |V_{ud} V_{cb} V_{cd}^*|
\{ 2 Re[ e^{i \gamma} (c^{ct}) (c^{ut})^*] {\cal X}
+ 8 Re[ e^{i \gamma} (\tilde c^{ct}) (\tilde c^{ut})^*] {\cal Y}  \nonumber \\
&+& 2 Re[ e^{i \gamma} (c^{ct}) (\tilde c^{ut})^*
+e^{-i \gamma} (c^{ut}) (\tilde c^{ct})^* ] {\cal Z} \}~ , \nonumber \\
{\cal M}_0 &=& 2 G_F^2 f_{\pi}^2 |V_{cb} V_{cd}^*|^2
\{ |c^{ct}|^2 {\cal X} + 4 |\tilde c^{ct}|^2 {\cal Y}
+2 Re[(c^{ct}) (\tilde c^{ct})^* ] {\cal Z} \}~,
\label{M2M1M0}
\end{eqnarray}
and
\begin{eqnarray}
{\cal X} &=& (m_b^2 +m_u^2) (p_b \cdot p_u) -2 m_b^2 m_u^2~, \nonumber \\
{\cal Y} &=& { m_{\pi}^4 \over (m_d +m_u)^2 } (p_b \cdot p_u)~, \nonumber \\
{\cal Z} &=& { 2 m_u m_{\pi}^2 \over m_d +m_u } ( p_b \cdot p_u - m_b^2 )~,
\nonumber \\
c^{ut} &=& \left( {c_1 \over N_c} + c_2 \right) - {1 \over N_c} (c_3^u -c_3^t)
- (c_4^u -c_4^t) - {1 \over N_c} (c_9^u -c_9^t) - (c_{10}^u -c_{10}^t)~ ,
\nonumber \\
c^{ct} &=& - {1 \over N_c} (c_3^c -c_3^t) - (c_4^c -c_4^t)
- {1 \over N_c} (c_9^c -c_9^t) - (c_{10}^c -c_{10}^t)~ , \nonumber \\
\tilde c^{qt} &=& {1 \over N_c} (c_5^q -c_5^t) +(c_6^q -c_6^t)
+{1 \over N_c} (c_7^q -c_7^t) +(c_8^q -c_8^t) \ \ \ (q = u, c)~ .
\end{eqnarray}
Here we have used the usual definition
of the phase angle,
\begin{eqnarray}
\gamma (\equiv \phi_3) = Arg[-(V_{ud} V^*_{ub})/(V_{cd} V^*_{cb})]~. \nonumber
\end{eqnarray}
\begin{figure}[htb]
\vspace{-3cm}
\begin{center}
\epsfig{file=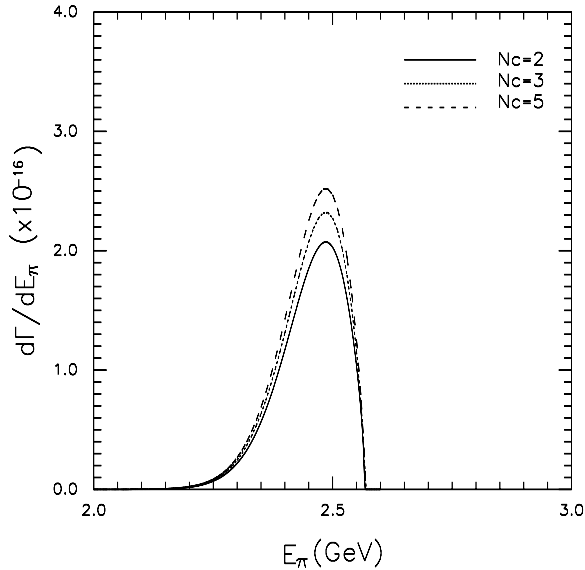,width=12cm}
\end{center}
\vspace{-4 cm}
\caption {$d\Gamma \over dE_{\pi}$ (in units of $10^{-16}$) versus $E_{\pi}$
for $\overline B^0 \to \pi^- X$ decay.  The solid, dotted, and dashed lines
correspond to $N_c = 2,~3,~5$, respectively.}
\vspace{0.5cm}
\end{figure}

\begin{figure}[htb]
\vspace{-3cm}
\begin{center}
\epsfig{file=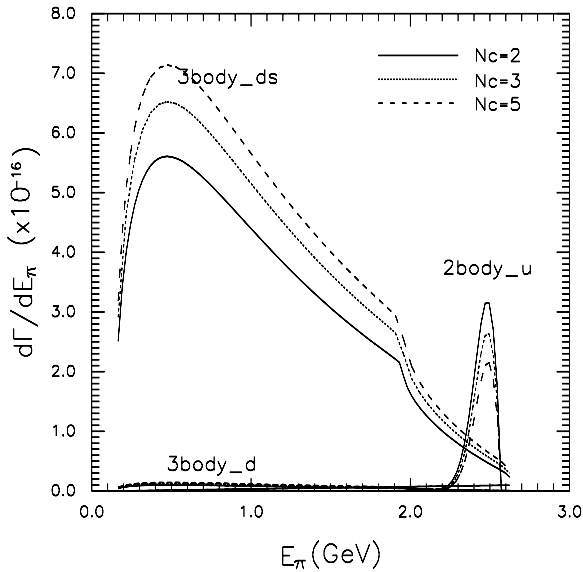,width=12cm}
\end{center}
\vspace{-4 cm}
\caption {$d\Gamma \over dE_{\pi}$ (in units of $10^{-16}$) versus $E_{\pi}$
for $B^- \to \pi^- X$ decay.  The solid, dotted, and dashed lines
correspond to $N_c = 2,~3,~5$, respectively.
Here, `2body\_u' stands for the two-body type ($b \to \pi^- u)$ processes
(see Fig. 2(a,c)), and
`3body\_ds' stands for the three-body type processes from the color-suppressed
tree and the $b \to d$ and $b \to s$ penguins (Fig. 2(b,d)),
while `3body\_d' stands for the three-body type process from
the $b \to d$ penguin only (Fig. 2(d)). }
\vspace{0.5cm}
\end{figure}

We first calculate the decay distribution in the $b$ quark rest frame and
boost it to the $B-$meson rest frame.
In the $b$ quark rest frame, the decay distribution is given by
\begin{eqnarray}
\left. {d \Gamma \over d E_{\pi}}\right|_b = {1 \over 16 \pi}
|{\cal M}|^2 {p_{\pi} \over m_b E_u} \delta (m_b - E_{\pi} - E_u )~.
\end{eqnarray}
In the $B-$meson rest frame, the $b$ quark is in motion and has the
energy $E_b$ satisfying the relation:
$E_b =m_B -E_u~$,
where $E_i = \sqrt{p^2 +m^2_i}$. $E_i$ and $m_i$ denote the energy and
the mass of $i$ quark inside
the $B-$meson, respectively.  The 3-momentum $p$ is defined by
$p = |\vec p_b| = |\vec p_u|$.
Thus, the $b$ quark mass is now a function of $p$ given by
\begin{eqnarray}
m_b^2 (p) = m_B^2 +m_u^2 -2 m_B \sqrt{p^2 +m_u^2}~.
\end{eqnarray}
The decay distribution in the $B$ rest frame can be calculated by
\begin{eqnarray}
{d \Gamma \over d E_{\pi}} = \int^{p_{max}}_0 dp \; p^2 \phi(p)
\left. {d \Gamma \over d E_{\pi}}\right|_b ~,
\end{eqnarray}
where $p_{max} = (m_B^2 -m_u^2)/ (2 m_B)$.
We have used the ACCMM model \cite{accmm} using the $B-$meson wave function:
\begin{eqnarray}
\phi (p) = {4 \over \sqrt{\pi p_{_F}^3}} e^{-p^2 / p_{_F}^2}~,
\end{eqnarray}
with the normalization $\int_0^{\infty} p^2 \phi(p) dp =1$.
\begin{table}
\caption{The branching ratio (BR) of $\overline B^0 \to \pi^- X$ for
fixed $\gamma=65^0$.  Here ${\cal B}$ denotes the BR calculated
for $|V_{ub}| = 0.004$.}
\smallskip
\begin{tabular}{c|cccc}
$N_c$ & ${\cal B} \times 10^{4}$ & ${\cal B}_2 \times 10^{4}$  &
${\cal B}_1 \times 10^{4}$ & ${\cal B}_0 \times 10^{4}$
\\ \hline 2 & 1.05 & 0.89 & 0.14 & 0.016
\\ 3 & 1.17 & 0.98 & 0.17 & 0.019
\\ 4 & 1.24 & 1.03 & 0.19 & 0.020
\\ 5 & 1.27 & 1.06 & 0.19 & 0.021
\end{tabular}
\end{table}
The Fermi momentum $p_{_F} = 0.3 {\rm GeV}$ has been used.
In Figs. 3 and 4 we show the decay distributions for $\overline B^0 \to \pi^- X$
and $B^- \to \pi^- X$ as a function of the charged pion energy $E_{\pi}$.
For $\overline B^0 \to \pi^- X$ decay, the distribution has a  peak at
$E_{\pi} \simeq m_B /2$.  This is a characteristic of a two-body decay.
Because the $b$ quark inside the $\overline B^0$ is in motion, the distribution has
some width shown in Fig. 3.
For $B^- \to \pi^- X$, the decay distribution is a mixture of three-body type
decay distributions and a two-body type decay distribution.
Figure 4 shows that the dominant
contribution arises from a three-body type decay (the $b \to s$ penguin process),
as explained in Sec. II.

The BR of $\overline B^0 \to \pi^- X$ can be expressed as
a polynomial of $|V_{ub}|$:
\begin{eqnarray}
{\cal B}(\overline B^0 \to \pi^- X)
    = \left|{V_{ub} \over 0.004} \right|^2 \cdot {\cal B}_2
    + \left|{V_{ub} \over 0.004}
    \right| \cdot {\cal B}_1 + {\cal B}_0 ~,
\label{BR}
\end{eqnarray}
where for convenience we have scaled $|V_{ub}|$ by the factor 0.004
(the central value of the OPAL data).
Tables I and II show the BR of $\overline B^0 \to \pi^- X$ for fixed
$\gamma (\equiv \phi_3) =65^\circ$ and $N_c =3$, respectively,
with a fixed input value of $|V_{ub}| = 0.004$.
We note that the term ${\cal B}_2$ is the dominant contribution
($\sim 10^{-4}$) to ${\cal B}$, while the contribution from the term
${\cal B}_0$ is very small ($\sim 10^{-6}$).
This is due to the fact (see Fig. 1) that ${\cal B}_2$
corresponds to mostly the tree contribution ($b \to u \bar u d$),
while ${\cal B}_0$ corresponds to the pure $b \to d$ penguin contribution
which is very small compared to the tree contribution,
and ${\cal B}_1$ corresponds to the interference between them.
We see that the BR of $\overline B^0 \to \pi^- X$ is about $10^{-4}$
for different values of $\gamma (\equiv \phi_3)$ and $N_c$.
\begin{figure}[htb]
\vspace{-3 cm}
\begin{center}
\epsfig{file=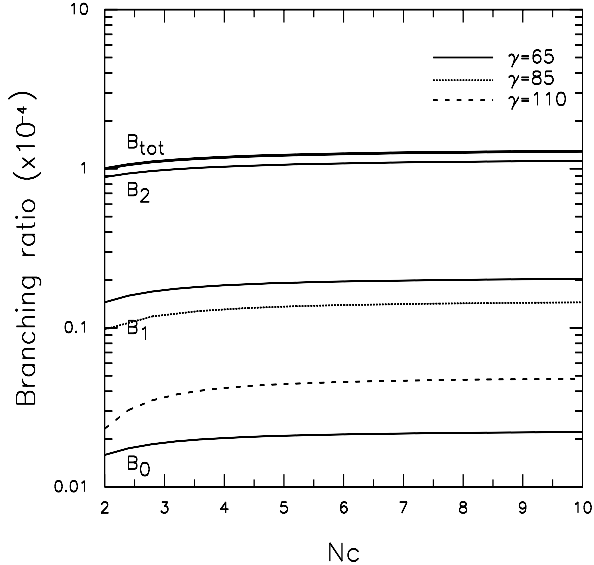,width=12cm}
\end{center}
\vspace{-4 cm}
\caption {The branching ratio (in units of $10^{-4}$) versus the effective number
of color, $N_c$,
for $\overline B^0 \to \pi^- X$ decay.
$B_{tot} (\equiv {\cal B})$ has been calculated
using $|V_{ub}| = 0.004$ and is denoted by the bold solid line.
The solid, dotted, and dashed lines
correspond to $\gamma = 65^\circ,~85^\circ,~110^\circ$, respectively.
For light quark masses, $m_u = 5$ MeV and $m_d = 7$ MeV have been used.}
\vspace{0.5cm}
\end{figure}
\begin{table}
\caption{The branching ratio of $\overline B^0 \to \pi^- X$ for fixed
$N_c =3$. Here ${\cal B}$ denotes the BR calculated for $|V_{ub}|
= 0.004$.}
\smallskip
\begin{tabular}{c|cccc}
$\gamma$ & ${\cal B} \times 10^{4}$  & ${\cal B}_2 \times 10^{4}$
& ${\cal B}_1 \times 10^{4}$ & ${\cal B}_0 \times 10^{4}$
\\ \hline 60 & 1.18 & 0.98 & 0.18 & 0.019
\\ 70 & 1.16 & 0.98 & 0.16 & 0.019
\\ 80 & 1.14 & 0.98 & 0.14 & 0.019
\\ 90 & 1.11 & 0.98 & 0.11 & 0.019
\\ 100 & 1.07 & 0.98 & 0.072 & 0.019
\\ 110 & 1.04 & 0.98 & 0.037 & 0.019
\end{tabular}
\end{table}
\begin{figure}[htb]
\vspace{-3 cm}
\begin{center}
\epsfig{file=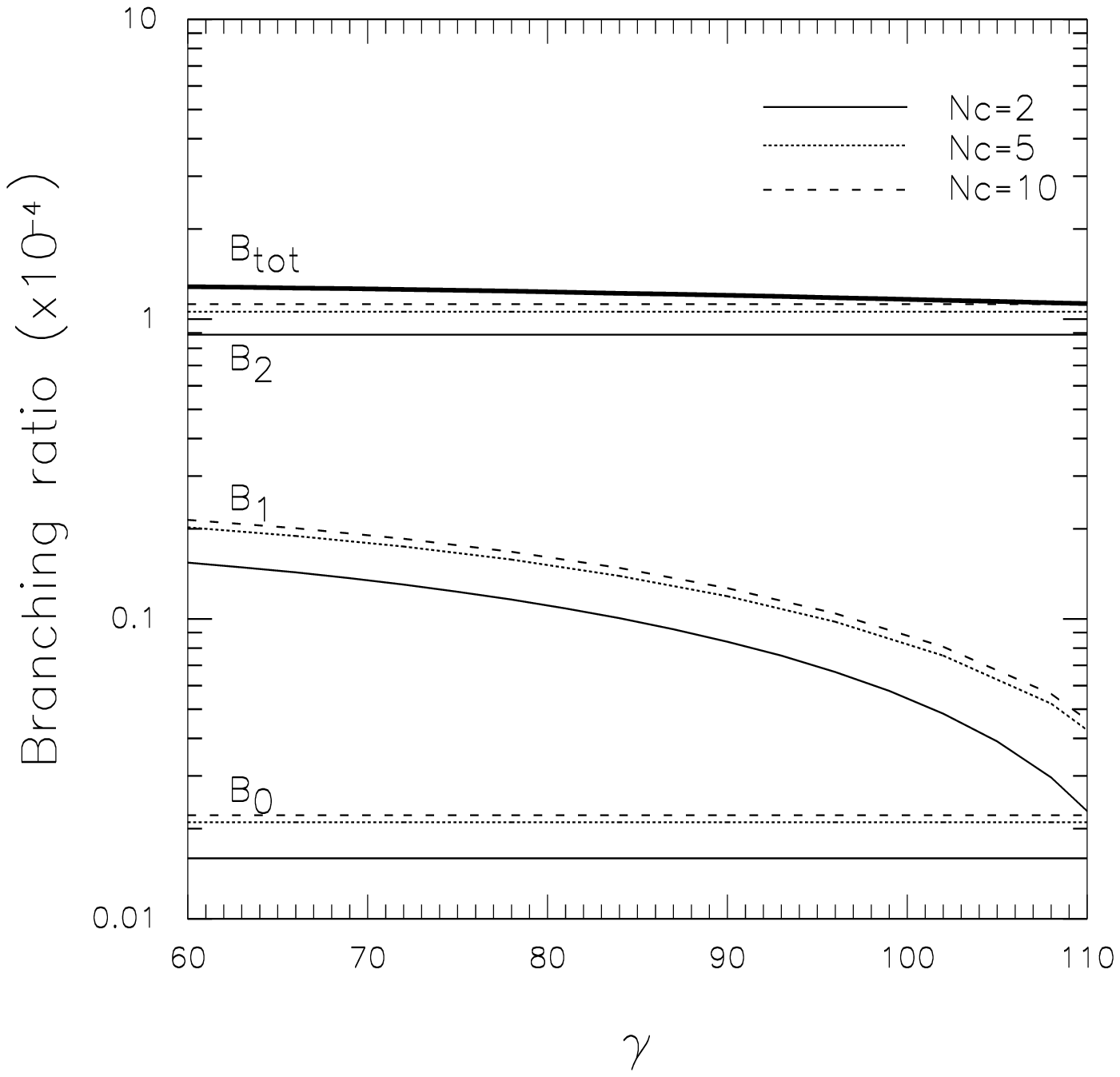,width=12cm}
\end{center}
\vspace{-4 cm}
\caption {The branching ratio (in units of $10^{-4}$) versus the CP phase, $\gamma$,
for $\overline B^0 \to \pi^- X$ decay. $B_{tot} (\equiv {\cal B})$
has been calculated
using $|V_{ub}| = 0.004$ and is denoted by the bold solid line.
The solid, dotted, and dashed lines
correspond to $N_c = 2,~5,~10$, respectively.
$m_u = 5$ MeV and $m_d = 7$ MeV have been used.}
\vspace{0.5cm}
\end{figure}

\begin{figure}[htb]
\vspace{-3 cm}
\begin{center}
\epsfig{file=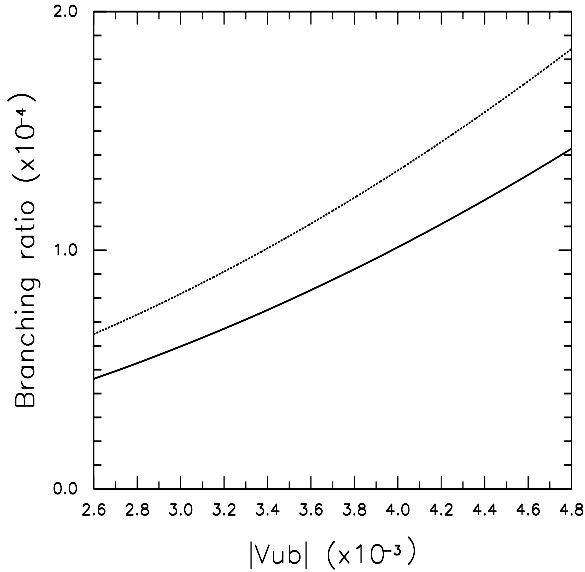,width=12cm}
\end{center}
\vspace{-4 cm}
\caption {The branching ratio (in units of $10^{-4}$) versus $|V_{ub}|$ for
$\overline B^0 \to \pi^- X$ decay.
The solid and the dotted lines correspond to the smallest and the largest
value of ${\cal B}$ in the given parameter space, respectively.
For light quark masses, $m_u = (1.5 - 5)$ MeV and $m_d = (3 - 9)$ MeV have been used.}
\vspace{0.5cm}
\end{figure}

In Fig. 5, we present the BR of $\overline B^0 \to \pi^- X$ as a function of
$N_c$ for three different values of $\gamma (\equiv \phi_3)=
65^\circ,~ 85^\circ,~ 110^\circ$.
As one can see from Eqs. (\ref{Msquared}, \ref{M2M1M0}, \ref{BR}) and
from Table 2 and Fig. 6,
${\cal B}_2$ and ${\cal B}_0$ are independent of $\gamma (\equiv \phi_3)$,
and only ${\cal B}_1$ depends on $\gamma (\equiv \phi_3)$.
Three different lines for ${\cal B}_1$ correspond to the relevant values
of $\gamma (\equiv \phi_3)$, respectively.
It is clearly shown that ${\cal B}_2$ is dominant.
A representative value of ${\cal B}$ for
$|V_{ub}| =0.004$ and $\gamma (\equiv \phi_3)= 85^\circ$ is shown as the bold
solid line in the figure.
The value of ${\cal B}$ does not vary much as $N_c$ varies.

Similarly, Figure 6 shows the BR of $\overline B^0 \to \pi^- X$ as a function
of $\gamma (\equiv \phi_3)$ for three different values of $N_c =2,~5,~10$.
The solid line corresponds to the case $N_c =2$, and the dotted line and
the dashed line are for $N_c =5$ and $N_c =10$, respectively.
The ${\cal B}_2$, ${\cal B}_1$ and ${\cal B}_0$ increase as
$N_c$ increases.  However, since the dominant term ${\cal B}_2$ does not
change much when  $N_c$ is varied, the BR
does not change much either.
In Fig. 6, a representative value of ${\cal B}$ for $|V_{ub}|=0.004$ and
$N_c =5$ is shown as the bold solid line.

Finally we summarize our result in Fig. 7.
The BR of $\overline B^0 \to \pi^- X$ is presented as a function of $|V_{ub}|$.
For light quark masses,
we use $m_u = (1.5 - 5)$ MeV and $m_d = (3 - 9)$ MeV.
We also vary the value of $N_c$ and $\gamma (\equiv \phi_3)$ in a reasonable
range: from $N_c =2$ to $10$, and from $\gamma (\equiv \phi_3)=60^\circ$
to $110^\circ$.
The solid and the dotted lines correspond to the smallest and the largest
value of ${\cal B}$ in the given parameter space, respectively.
For the given $|V_{ub}|$, the BR  is estimated with a relatively
small error ($<20 \%$), as can be seen.
Reversely, for the given (i.e., experimentally measured) BR,
the value of $|V_{ub}|$ can be determined with a reasonably small error
($\sim 15 \%$).
(Of course, since in practice the BR would be measured
with some errors, $|V_{ub}|$ could be determined with larger error:
e.g., for ${\cal B} = (1.0 \pm 0.1) \times 10^{-4}$, our result suggests
$|V_{ub}| =\simeq (3.7 \pm 0.59) \times 10^{-3}$.)

\begin{table}
\caption{The branching ratio of $B^0 \to \pi^- X$ for fixed
$\gamma =65^\circ$ and $N_c =3$.
Here ${\cal B}$ denotes the BR calculated for $|V_{ub}|= 0.004$
and $2.32 {\rm GeV} < E_{\pi} < 2.56 {\rm GeV}$. }
\smallskip
\begin{tabular}{ccc||cc}
$\gamma = 65^\circ$ & & & $N_c =3$ & \\
$N_c$ & ${\cal B} \times 10^{4}$ & & $\gamma$
& ${\cal B} \times 10^{4}$
\\ \hline 2 & 0.95 & & 60 & 1.06
\\ 3 & 1.05 & & 80 & 1.04
\\ 4 & 1.11 & & 100 & 1.01
\\ 5 & 1.14 & & 110 & 0.99
\end{tabular}
\end{table}
In order to use the decay process $\overline B^0 \to \pi^- X$,
one may need to consider the
$B^0 - \overline B^0$ mixing effect: $\overline B^0 \to B^0 \to \pi^- X$.
The neutral $\overline B^0$ has about $18\%$ probability of decaying as the
opposite flavor $B^0$ \cite{rosner}.
Thus, including the $B^0 - \overline B^0$ mixing effect, the decay rate $\Gamma_0$
for $\overline B^0$ decay to $\pi^- X$ can be expressed as
\begin{eqnarray}
\Gamma_0 = (0.82)\cdot \Gamma(\overline B^0 \to \pi^- X) + (0.18)\cdot
\Gamma(B^0 \to \pi^- X)~,
\label{Gamma}
\end{eqnarray}
where $\Gamma(\overline B^0 (B^0) \to \pi^- X)$ denotes the decay rate for
$\overline B^0 (B^0)$ decay {\it directly} to $\pi^- X$.
The BR of $B^0$ decay {\it directly} to $\pi^- X$ is about $90\%$
of the BR of $\overline B^0$ decay {\it directly} to $\pi^- X$
with the energy cut\footnote
{As mentioned in Sec. II,
the charged pion in the decay mode $B^0 \to \pi^- X$
contains the spectator quark $d$, and
this process is basically a three-body decay
$B^0 \to \pi^- u \bar d ~(\bar s)$.
Therefore, in order to remove this large $s$ quark contribution, one needs to make
such a large energy cut  (see Fig. 4).},
$2.32 {\rm GeV} < E_{\pi} < 2.56 {\rm GeV}$,
as can be seen from Table III.
Even though the theoretical estimate for the mode $B^0 \to \pi^- X$
would include a somewhat larger uncertainty, which mainly arises from
the relevant hadronic form factor, the total
error of $\Gamma_0$ would not increase much.  For example, if the estimate of
$\Gamma(B^0 \to \pi^- X)$ in Eq. (\ref{Gamma}) includes an error of $30\%$,
then its actual contribution to the final error of $\Gamma_0$ is less than
$5\%$.
Therefore, even after considering the effect from the $B^0 - \overline B^0$ mixing,
our result holds with reasonable accuracy.

\section{Conclusions}

We have studied semi-inclusive charmless decays of $B-$mesons to $\pi X$
in the final state, such as $\overline B^0 \to \pi^{\pm (0)} X$,
$B^0 \to \pi^{\pm (0)} X$, $B^{\pm} \to \pi^{\pm (0)} X$,
where $X$ does not contain a charm (anti)quark.
Among these $B \to \pi X$ decays, we have found that the mode
$\overline B^0 \to \pi^- X$ ($B^0 \to \pi^+ X$) is particularly
interesting and can be used to determine of the CKM matrix element
$|V_{ub}|$ in phenomenological studies.

In $\overline B^0 \to \pi^- X$  decay,
the charged pion  in the final state
can be produced via $W-$boson emission at tree level and is expected to be
energetic ($E_{\pi} \simeq m_B / 2$).
Thus, the energetic charged pion   in the final
state can be a characteristic signal for this mode.
This process is basically a \emph{two}-body
decay process of $b \to \pi^- u$.
As a result, in this mode,
the model-dependence does not appear to be severe.

We have calculated the BR of $\overline B^0 \to \pi^- X$ and presented it as
a function of $|V_{ub}|$.
It is expected that its BR is an order of $10^{-4}$.
(In this analysis, higher-order QCD corrections have not been considered;
instead we analyzed within the QCD improved general factorization framework. So,
a further study on this process would be very interesting.)
We have also estimated the possible
uncertainty due to $B^0 - \overline B^0$ mixing effects via the decay chain
$\overline B^0 \to B^0 \to \pi^- X$.
Other theoretical uncertainties, such as those arising from the WC's
and the CKM elements, could affect our results in some extent.
However, as soon as the relevant results from the experiments become available, one
can use them to reduce theoretical uncertainties in turn.  Thus, in the viewpoint of
phenomenological studies, our results can be used to determine $|V_{ub}|$
with reasonable accuracy by measuring the BR of $\overline B^0 \to \pi^- X$.
Therefore, the process $\overline B^0 \to \pi^- X$
($B^0 \to \pi^+ X$) can play an important role
in measuring $|V_{ub}|$ at $B-$factories.
\\

\centerline{\bf ACKNOWLEDGEMENTS}
\medskip

\noindent We thank F. Krueger and Y. Kwon for their valuable comments.
The work of C.S.K was supported by
Grant No. 2001-042-D00022 of the KRF.
The work of J.L. was supported by Grant No. R03-2001-00010 of the KOSEF.
The work of S.O was supported
by  CHEP-SRC Program, Grant No. 20015-111-02-2
and Grant No. R02-2002-000-00168-0 from BRP of the KOSEF.
The work of D.Y.K., H.S.K, and J.S.H was
supported by the BSRI Program of MOE, Project No. 99-015-D10032.


\end{document}